\input harvmac
\noblackbox
 
\font\ticp=cmcsc10
 
\def\Title#1#2{\rightline{#1}\ifx\answ\bigans\nopagenumbers\pageno0\vskip1in
\else\pageno1\vskip.8in\fi \centerline{\titlefont #2}\vskip .5in}

\font\ticp=cmcsc10
\font\ttsmall=cmtt10 at 8pt

\input epsf
\ifx\epsfbox\UnDeFiNeD\message{(NO epsf.tex, FIGURES WILL BE
IGNORED)}
\def\figin#1{\vskip2in}
\else\message{(FIGURES WILL BE INCLUDED)}\def\figin#1{#1}\fi
\def\ifig#1#2#3{\xdef#1{fig.~\the\figno}
\goodbreak\topinsert\figin{\centerline{#3}}%
\smallskip\centerline{\vbox{\baselineskip12pt
\advance\hsize by -1truein\noindent{\bf Fig.~\the\figno:} #2}}
\bigskip\endinsert\global\advance\figno by1}

%
%
\def\[{\left [}
\def\]{\right ]}
\def\({\left (}
\def\){\right )}
\def\p{\partial}
\def\R{{\bf R}}
\def\S{{\bf S}}
\def\l{\ell}

\def\C{{\cal{C}}}

\def\e{\varepsilon}
\def\Om{\Omega}
\def\a{\alpha}

\def\CN{{\cal N}}

\def\CV{{\cal V}}
\def\CM{{\cal M}}

\def\dua{\( {\p \over \p u} \)^a}
\def\dva{\( {\p \over \p v} \)^a}

\def\ie{{\it i.e.}}

\lref\bmn{
D.~Berenstein, J.~M.~Maldacena and H.~Nastase,
{\it Strings in flat space and pp waves from N = 4 super Yang Mills},
JHEP {\bf 0204}, 013 (2002)
[arXiv:hep-th/0202021].}

\lref\bfhp{
M.~Blau, J.~Figueroa-O'Farrill, C.~Hull and G.~Papadopoulos,
{\it A new maximally supersymmetric background of IIB superstring theory},
JHEP {\bf 0201}, 047 (2002)
[arXiv:hep-th/0110242].}

\lref\metsaev{
R.~R.~Metsaev,
{\it Type IIB Green-Schwarz superstring in plane wave Ramond-Ramond 
background},
Nucl.\ Phys.\ B {\bf 625}, 70 (2002)
[arXiv:hep-th/0112044].}

\lref\nobh{
V.~E.~Hubeny and M.~Rangamani,
{\it No horizons in pp-waves},
[arXiv:hep-th/0210234].}

\lref\causal{
V.~E.~Hubeny and M.~Rangamani,
{\it Causal structures of pp-waves},
[arXiv:hep-th/0211195].}

\lref\maro{
D.~Marolf and S.~F.~Ross,
{\it Plane waves: To infinity and beyond!},
[arXiv:hep-th/0208197].}

\lref\penrose{
R.~Penrose,
``Any spacetime has a planewave as a limit,'' in
{\it Differential geometry and relativity}, pp 271-275,
Reidel, Dordrecht, 1976.}

\lref\penr{
R.~Penrose,
{\it A Remarkable Property Of Plane Waves In General Relativity},
Rev.\ Mod.\ Phys.\  {\bf 37}, 215 (1965).}

\lref\gava{
D.~Garfinkle and T.~Vachaspati,
{\it Cosmic String Traveling Waves},
Phys.\ Rev.\ D {\bf 42}, 1960 (1990).}

\lref\garfinkle{
D.~Garfinkle,
{\it Black String Traveling Waves},
Phys.\ Rev.\ D {\bf 46}, 4286 (1992)
[arXiv:gr-qc/9209002].}

\lref\bena{
D.~Berenstein and H.~Nastase,
{\it On lightcone string field theory from super Yang-Mills and holography},
[arXiv:hep-th/0205048].}

\lref\kmr{
N.~Kaloper, R.~C.~Myers and H.~Roussel,
{\it Wavy strings: Black or bright?},
Phys.\ Rev.\ D {\bf 55}, 7625 (1997)
[arXiv:hep-th/9612248].}

\lref\gaunt{
J.~P.~Gauntlett, J.~B.~Gutowski, C.~M.~Hull, S.~Pakis and H.~S.~Reall,
{\it All supersymmetric solutions of minimal supergravity in five 
dimensions},
[arXiv:hep-th/0209114].}

\lref\zv{
L.~A.~Pando Zayas and D.~Vaman,
{\it Strings in RR plane wave background at finite temperature},
[arXiv:hep-th/0208066].}

\lref\gss{
B.~R.~Greene, K.~Schalm and G.~Shiu,
{\it On the Hagedorn behaviour of pp-wave strings and N = 4 SYM 
theory at  finite R-charge density},
[arXiv:hep-th/0208163].}

\lref\sugawara{
Y.~Sugawara,
{\it Thermal Amplitudes in DLCQ Superstrings on PP-Waves},
[arXiv:hep-th/0209145].}

\lref\ghk{
S.~S.~Gubser, A.~Hashimoto, I.~R.~Klebanov and M.~Krasnitz,
{\it Scalar absorption and the breaking of the world volume conformal 
invariance},
Nucl.\ Phys.\ B {\bf 526}, 393 (1998)
[arXiv:hep-th/9803023].}

\lref\gubser{
S.~S.~Gubser and A.~Hashimoto,
{\it Exact absorption probabilities for the D3-brane},
Commun.\ Math.\ Phys.\  {\bf 203}, 325 (1999)
[arXiv:hep-th/9805140].}

\lref\ejp{
N.~Evans, C.~V.~Johnson and M.~Petrini,
{\it Clearing the throat: Irrelevant operators and finite temperature in  
large N gauge theory},
JHEP {\bf 0205}, 002 (2002)
[arXiv:hep-th/0112058].}

\lref\costa{
M.~S.~Costa,
{\it Absorption by double-centered D3-branes and the Coulomb branch of 
N = 4  SYM theory},
JHEP {\bf 0005}, 041 (2000)
[arXiv:hep-th/9912073].}

\lref\costab{
M.~S.~Costa,
{\it A test of the AdS/CFT duality on the Coulomb branch},
Phys.\ Lett.\ B {\bf 482}, 287 (2000)
[Erratum-ibid.\ B {\bf 489}, 439 (2000)]
[arXiv:hep-th/0003289].}

\lref\dgks{
U.~H.~Danielsson, A.~Guijosa, M.~Kruczenski and B.~Sundborg,
{\it D3-brane holography},
JHEP {\bf 0005}, 028 (2000)
[arXiv:hep-th/0004187].}

\lref\dast{
S.~R.~Das, S.~Kalyana Rama and S.~P.~Trivedi,
{\it Supergravity with self-dual B fields and instantons in 
noncommutative  gauge theory},
JHEP {\bf 0003}, 004 (2000)
[arXiv:hep-th/9911137].}

\lref\witten{
E.~Witten,
{\it Anti-de Sitter space, thermal phase transition, and 
confinement in  gauge theories},
Adv.\ Theor.\ Math.\ Phys.\  {\bf 2}, 505 (1998)
[arXiv:hep-th/9803131].}

\lref\bcr{
D.~Brecher, A.~Chamblin and H.~S.~Reall,
{\it AdS/CFT in the infinite momentum frame},
Nucl.\ Phys.\ B {\bf 607}, 155 (2001)
[arXiv:hep-th/0012076].
}

\lref\gy{
G.~T.~Horowitz and H.~s.~Yang,
{\it Black strings and classical hair},
Phys.\ Rev.\ D {\bf 55}, 7618 (1997)
[arXiv:hep-th/9701077].
}

\lref\myers{
R.~C.~Myers,
{\it Pure states don't wear black},
Gen.\ Rel.\ Grav.\  {\bf 29}, 1217 (1997)
[arXiv:gr-qc/9705065].
}

\lref\ross{
S.~F.~Ross,
{\it Singularities in wavy strings},
JHEP {\bf 9808}, 003 (1998)
[arXiv:hep-th/9710158].
}

\lref\bmz{
P.~Bain, P.~Meessen and M.~Zamaklar,
{\it Supergravity solutions for D-branes in Hpp-wave backgrounds},
[arXiv:hep-th/0205106].
}

\lref\ct{
M.~Cvetic and A.~A.~Tseytlin,
{\it Solitonic strings and BPS saturated dyonic black holes},
Phys.\ Rev.\ D {\bf 53}, 5619 (1996)
[Erratum-ibid.\ D {\bf 55}, 3907 (1997)]
[arXiv:hep-th/9512031].
}

\lref\TseytlinA{
A.~A.~Tseytlin,
{\it Extreme dyonic black holes in string theory},
Mod.\ Phys.\ Lett.\ A {\bf 11}, 689 (1996)
[arXiv:hep-th/9601177].
}

\lref\TseytlinB{
A.~A.~Tseytlin,
{\it Composite BPS configurations of p-branes in 10 and 11 dimensions},
Class.\ Quant.\ Grav.\  {\bf 14}, 2085 (1997)
[arXiv:hep-th/9702163].
}

\lref\mone{
H.~W.~Lee, Y.~S.~Myung, J.~Y.~Kim and D.~K.~Park,
{\it Propagating waves in extremal black string},
Mod.\ Phys.\ Lett.\ A {\bf 12}, 545 (1997)
[arXiv:hep-th/9607001].
}

\lref\mtwo{
H.~W.~Lee, Y.~S.~Myung, J.~Y.~Kim and D.~K.~Park,
{\it Higher dimensional extremal black strings},
[arXiv:hep-th/9610031].
}

%
\baselineskip 16pt
\Title{\vbox{\baselineskip12pt
\line{\hfil SU-ITP-02/44}
\line{\hfil UCB-PTH-02/52}
\line{\hfil LBNL-51739}
\line{\hfil \tt hep-th/0211206} }}
{\vbox{
{\centerline{}Generating asymptotically plane wave spacetimes
}}}
\centerline{\ticp Veronika E. Hubeny$^a$
 and Mukund Rangamani$^{b,c}$ \footnote{}{\ttsmall
 veronika@itp.stanford.edu, mukund@socrates.berkeley.edu}}
\bigskip
\centerline {\it $^a$
Department of Physics, Stanford University, Stanford, CA 94305, USA} 
\centerline{\it $^b$ Department of Physics, University of California,
Berkeley, CA 94720, USA} 
\centerline{\it $^c$ Theoretical Physics Group, LBNL, Berkeley, CA 94720, USA}

\bigskip
\centerline{\bf Abstract}
\bigskip

\noindent
In an attempt to study asymptotically plane wave spacetimes which 
admit an event horizon,  
we find solutions to vacuum Einstein's equations in arbitrary
dimension which have a globally null Killing field and rotational symmetry.
We show that while such solutions can be deformed to 
include ones which are asymptotically 
plane wave, they do not posses a regular event horizon. If we allow for 
additional matter, such as in supergravity theories, we show that it is 
possible to have extremal solutions with globally null Killing field, a 
regular horizon, and which, in addition, are asymptotically plane wave. 
In particular, we deform the extremal M2-brane solution in 11-dimensional 
supergravity so that it behaves asymptotically as a 10-dimensional 
vacuum plane wave times a real line.
 
\Date{November, 2002}
%

\newsec{Introduction}

Plane wave spacetimes have been on the forefront of attention over the 
last few months, stemming from the observation that the effective dynamics
of a certain sector of the ${\cal{ N} }  = 4 $, $ d= 4$ supersymmetric 
Yang-Mills theory with large R-charge, is encoded in the dynamics of 
strings propagating in the maximally supersymmetric plane wave 
background of Type IIB supergravity \bmn. It is even more fascinating that 
the world-sheet sigma model is a solvable conformal field theory, despite 
there being non-trivial Ramond-Ramond fluxes in the background \bmn,\metsaev. 

An interesting variation of the BMN correspondence is to 
consider a thermal version of the same ({\it cf.} \zv, \gss, \sugawara,
for computations of the partition function). 
Naively, one would imagine that the 
aforementioned gauge theory system in a thermal ensemble ought to have its 
dynamics encoded in the propagation of strings in an asymptotically 
plane wave black hole background. This would be in direct analogy with  
the standard story in the AdS/CFT correspondence \witten, wherein 
thermalizing the gauge theory corresponds to looking at black holes 
in the dual supergravity background.
From this standpoint it would be very interesting to examine whether 
there can be black holes which are asymptotically plane wave. 

Although plane waves have been known as solutions to Einstein's equations for 
many decades now, very little is known about these spacetimes. 
While a few interesting facets of information, 
such as the fact that the Penrose limit \penrose\ of any spacetime 
is a plane wave, and some global properties such as these spacetimes 
not being globally hyperbolic \penr, have been known for a while, 
many questions remain unexplored. In particular, 
it is only recently that the causal structures of some 
plane wave spacetimes were constructed \bena, \maro, \causal.

In a previous paper \nobh, we had asked whether spacetimes 
which are of the pp-wave form can admit event horizons. We gave evidence 
that this is not possible, by demonstrating the possibility of causal 
communication from any point in the spacetime manifold out to ``infinity''. 
We have subsequently re-confirmed the absence of horizons
 \causal\ by considering the full causal structure of pp-waves.
This basically implies that spacetimes admitting a covariantly
constant null Killing field do not admit event horizons.

A natural question which then presents itself is whether one can 
find black hole solutions by relaxing some part of the 
requirement of a covariantly constant null Killing field.
As a second step, we demonstrated in \nobh\ that there exist spacetimes which 
admit a globally null, Killing, but not covariantly constant, vector 
field, which have a regular horizon, and in addition are asymptotically 
plane wave. The strategy was to start with an asymptotically flat 
spacetime with a globally null, hypersurface-orthogonal Killing field 
which admits a regular horizon, and deform it to be  asymptotically plane wave
by using the Garfinkle-Vachaspati method \gava, \garfinkle. In particular, 
we exhibited a five-dimensional charged black string solution \kmr\
which had the requisite properties, and deformed it to be asymptotically 
plane wave. 

We would like to enquire whether there are analogous neutral black string 
solutions which may be asymptotically plane wave. We saw in \nobh\ that 
one cannot simply ``uncharge'' the solution
by modifying the parameters, because then the horizon shrinks
to zero size and becomes singular.  
So, instead, we explore whether we can actually find solutions
to vacuum Einstein's equations by directly solving them for a
general metric ansatz with the appropriate symmetries.

The general question we pose is: {\it Do there exist neutral 
black string solutions in vacuum gravity which admit a globally null Killing 
field and are asymptotically plane wave?} 
Making the additional requirement 
that the solution be rotationally symmetric, we find that such solutions 
do {\it not} exist. In particular, we find all solutions to 
vacuum Einstein's equations, with a globally defined null Killing 
field and rotational symmetry in the transverse plane. 
Although we can subsequently deform these to break the rotational symmetry
and induce the correct asymptotics, none of such solutions
can correspond to a black string, \ie, they don't admit horizons.

We find that there are two classes of solutions: one, which we shall
discuss below in some detail, which is asymptotically flat;
and the other, described by pure pp-waves.
The latter, as we have demonstrated \nobh, does not admit horizons, 
{\it i.e.}, it cannot represent the black string, albeit having the
correct asymptopia in case of plane waves.
On the other hand, the former is asymptotically flat; but it can be 
deformed to be asymptotically plane wave using the Garfinkle-Vachaspati 
method \gava, \garfinkle. However, these solutions have naked singularities,
so unless we can resolve the singularity and cloak it with an appropriate 
event horizon, they would not correspond to the solutions we are looking for.

This is a quite a remarkable result.  Restricting ourselves to 
vacuum Einstein's equations severely restricts the nature of the
possible solutions, and in particular eliminates the interesting
black string ones.
Thus, to obtain a neutral black object in asymptotically plane wave
spacetime, we might be forced to forgo some of the symmetries we
imposed above.  In particular, if there is no globally null Killing
field, one can imagine patching the ordinary vacuum black hole/string
into the plane wave.  We leave this for future consideration.

Given the absence of solutions to 
vacuum Einstein's equations with a globally null Killing field and a 
regular horizon, we might ask about the corresponding status in 
supergravity. There  the story is much richer and more 
promising. An interesting 
class of solutions corresponds to the extremal black brane solutions in 
Type II supergravity in ten dimensions or in eleven dimensional 
supergravity. As we will explicitly demonstrate below,
all the D$p$ branes with $p \ge 1$ and the M-brane solutions 
can be easily deformed to be asymptotically plane wave. The most 
interesting case is that of the M2-brane, since it is the only 
solution of the above class which admits a regular horizon with a 
singularity behind. This solution may be trivially deformed to one 
which is asymptotically a ten dimensional vacuum plane wave\foot{In 
what follows, we will denote the vacuum plane wave in $d$-dimensions 
as $\CV_d$} times 
a real line {\it i.e.}, $\CV_{10} \times \R$.

The organization of the paper is as follows. We begin in Section 2
with a discussion of the allowed solutions to vacuum 
Einstein's equations with the requisite 
symmetries. These will be the new solutions which we present in this work.
For clarity of presentation, we will first discuss 
solutions which have an additional  isometry and then show how 
one might consider more general solutions. After a brief discussion of 
their properties, we turn to a brief review of the Garfinkle-Vachaspati 
technique in Section 3, and demonstrate how it may be used to convert our 
vacuum solutions to have plane wave asymptotics. 
In Section 4, we turn to supersymmetric solutions and show how to 
construct interesting spacetimes which have an asymptotic plane wave 
structure. We end with a discussion in Section 5.

\newsec{Vacuum solutions}

In the present section, we will write down all solutions of 
vacuum Einstein's equations with a globally null Killing field
and rotational symmetry. The solutions we find fall into 
 two classes: 

\noindent
{\bf a.}
 pp-waves, which have a covariantly constant null Killing field.

\noindent
{\bf b.}
 Solutions with a null Killing field which is not covariantly 
constant. These solutions are singular and  asymptotically flat.

In order to simplify the discussion, we begin by demanding an additional 
 isometry. As we shall show later, it is possible to write down 
more general solutions by relaxing this requirement. 

\subsec{Ansatz}

We first write down the metric ansatz for the spacetime admitting
all the required symmetries. A naive guess would be to write down 
an ansatz for the metric motivated by the form of the 
extremal black string solution presented in \nobh,  
\ct, \TseytlinA, \kmr. The solution 
was given as 
\eqn\fdbh{
ds_{str}^2 = {2 \over h(r)}  \, du\, dv + {f(r) \over h(r)}
 \, du^2 + k(r) \, l(r) \(
dr^2 + r^2 d\Omega_2^2 \), }
which describes the metric in the string frame. 
However, as the dilaton depends 
only on the radial coordinate, the form of the 
solution remains the same in the Einstein frame. 
So we can write our metric ansatz as\foot{
In the following, we redefine $v \to -v$, to keep the notation
consistent with \causal\ and the bulk of \nobh.}
\eqn\ansatz{
ds^2 = {1 \over H(r)} \( -2 \, du \, dv + F(r) \, du^2 \)
      + G(r) \, (dr^2 + r^2 \, d\Om_{d-3}^2).}
In this section we will concern ourselves with purely the Einstein-Hilbert 
action with no matter. Given the absence of massless scalar fields,
we have no ambiguity in the choice of frame, {\it i.e.}, the above metric is 
the Einstein frame metric. 

This is in fact, up to diffeomorphisms,
the most general form of a metric satisfying the following 
properties:
it has a globally null Killing field, $\dva$, apart from 
possessing the additional Killing field $\dua$, and 
being spherically symmetric in the plane transverse to these 
two directions. The metric on these spheres given by $d\Om_{d-3}^2$, and 
the remaining radial coordinate is denoted by $r$.

In order to see that this is indeed the minimal ansatz with the 
requisite symmetry, realize that we have effectively a four-dimensional 
metric in the coordinates $(u,v,r,\Om)$, having fixed the metric on the 
spheres. This implies ten components of the metric. However, using the 
fact that the metric has a null symmetry and a rotational invariance, we 
may set $g_{vv}$ and $g_{i \Om}$ to zero for $i = (u,v,r)$. We might also 
fix the radial coordinate, so that the spheres of radius $r$ have the
appropriate area.  
This implies five independent functions and the metric takes the form
\eqn\oldansatz{
ds^2 = -2\, A(r) \,  du \, dv  + B(r) \, du^2 + 2 \, C(r)\, dv \, dr + 
2 \, D(r) \, du \, dr + E(r) \, dr^2 + r^2 \, d\Omega_{d-3}^2.
}
One can now define new coordinates by first choosing
$d\tilde{u} = du - {C(r) \over A(r) } \, dr$ 
and then $\tilde{v}$, satisfying 
$d\tilde{v} = dv - {1\over A(r)} \, \(D(r) + 
{B(r) \, C(r) \over A(r)} \) \, dr$. Then the metric reduces to the 
canonical form given in \ansatz, after some trivial redefinition 
of coordinates.

\subsec{General solution}

Solving the vacuum Einstein's equations in $d$ dimensions
with the ansatz \ansatz\
yields 3 classes of solutions.
First, we have the solution
\eqn\trivsoln{
G(r) = G_0 = {\rm const}, \ \ \ \ \ H(r)  = H_0 = {\rm const,} 
}
and $F(r)$ is harmonic in the transverse coordinate, namely
$F(r) = F_0 + {F_1 \over r^{d-4}}$ where $F_0$ and $F_1$ are constants.
But these are just the usual vacuum pp-waves.  We can see this explicitly
by fixing $H_0 \equiv 1$ by appropriate constant rescaling 
of $u$ and $v$, and similarly $G_0 \equiv 1$ by rescaling $r$.
The metric \ansatz\ then becomes
\eqn\pp{
ds^2 =  -2 \, du \, dv + F(r) \, du^2 + dr^2 + r^2 \, d\Om^2}
which is just the pp-wave metric.

The second class looks slightly less trivial, 
\eqn\ppsoln{
G(r) = {\a \over r^4}, \ \ \ \ \ H(r) = H_0 = {\rm const,} 
}
and again $F(r)$ is harmonic.
But these are in fact identical to the previous solutions \trivsoln.  
To see that, consider the coordinate transformation 
$r \to {1 \over r}$. 
Fixing the constants appropriately, we obtain \pp.

Finally, the third and most interesting class of solutions
in $d>4$ dimensions  is given by
\eqn\solng{
ds^2 = {1 \over H(r)} \( -2 \, du \, dv + F(r) \, du^2 \)
      + G(r) \, (dr^2 + r^2 \, d\Om_{d-3}^2)}

\eqn\solnH{
H(r) = 
 c_1 \( {r^{d-4} - a \over r^{d-4} + a} \)^
{\sqrt{{2 (d-3) \over (d-2)}}}
}

\eqn\solnG{
G(r) = 
 c_2 \( {H(r)^2 \over H'(r) \, r^{d-3} } \)^{2 \over d-4} }

\eqn\solnF{
F(r) = c_3  + c_4  \ln H(r),}
where $a$, $c_1$, $c_2$, $c_3$, and $c_4$ are arbitrary constants.
For the metric to have the correct signature, we require that $c_1>0$.
In fact, we can fix many of the constants to be $\equiv \pm 1$ 
 by appropriate diffeomorphisms.
For example,
we can fix $c_3 \equiv 1$ by rescaling $u$ and absorbing $c_3$   
in $c_4$ and $c_1$, 
then $c_1 \equiv 1$ by rescaling $u$ and $v$, and $c_2 \equiv 1$ by rescaling
$r$ and $a$.
Whether $\dua$ is a timelike or spacelike (or null) Killing field
depends on the sign on $F(r)/H(r)$, in particular, on the constants
involved.  For $c_1>0$, if both $c_4$ and $c_3$ in 
\solng\ have the same sign, 
then $\dua$ is null at some $r$, whereas if $c_3<0<c_4$,  $\dua$ is 
globally timelike and if $c_4<0<c_3$, it is globally spacelike.
(If $c_4=0$, we in fact have two globally null Killing fields.)
We can rewrite $G$ a bit more explicitly (with an appropriate choice 
of $c_2$ to absorb the $d$-dependent coefficient) as
\eqn\Gexpl{
G(r) = 
 \( H(r) \, {(r^{d-4} - a)(r^{d-4} + a) \over r^{2d-8} } \)^{2 \over d-4}
}
The solution in the special case of four dimensions is given as 
\eqn\fourdim{
ds_4^2 = \( 1 + \ln r \) \, \[ -2\, du \, dv + \( 1 + \ln \(1 + \ln r \)\) \, 
du^2 \] + {1 \over r^2 \sqrt{1 + \ln r }} \, \( dr^2 + r^2 d\theta^2 \).}
In writing the above we have made a particular choice of the 
constants of integration so as to present the solution in the simplest form.
\subsec{Properties of the solution}

Let us now focus on the last class of solutions, 
given by \solng, with \solnH, \solnG, and \solnF.
We see that these are asymptotically flat spacetimes, 
since we can fix $c_1 = c_2 = c_3 \equiv 1$, and then
$H(r) \to 1$ as $r \to \infty$, so that from \Gexpl,
$G(r) \to 1$ as $r \to \infty$, and clearly $F(r) \to 1$ as $r \to \infty$.

Also, these are explicitly distinct from pp-waves,
since  not all the curvature invariants vanish.
In fact, the curvature invariant composed from the Weyl tensor
blows up as $r^{d-4} \to \pm a$ ($r$ is positive but $a$ can have 
either sign).  This is suggestive of a curvature singularity;
one can in fact easily verify that these spacetimes are
singular by noting that radially in-going geodesics end there
at finite affine parameter.
Thus, we restrict our spacetime to $r > |a|^{1 \over d-4}$.

Furthermore, from the form of the metric, in particular from \solnH\ and
\solnG, it seems that there are no event horizons.
A quick way to 
verify the absence of horizons is to show that there are 
causal curves from any point of the spacetime which can communicate 
`out to infinity'. 
As we noted in \nobh, for any solution with a null Killing field $\dva$,
the orbits of the Killing field actually describe null geodesics;
so we can always reach $v \to \infty$ by a null geodesic. However, this
does not suffice, since we also want to show causal communication 
with large transverse directions.  To this end, we will present a 
causal (in fact a null) curve which reaches $r \to \infty$.
Consider in the metric \solng\ a null curve $\C$ 
parameterized by $\lambda$, with the coordinates along $\C$ satisfying
\eqn\nullcurve{\eqalign{
u(\lambda)& = \lambda \cr
v(\lambda) &= \a \, \lambda \cr
{d\Om \over d \lambda} & = 0 \cr
\({dr \over d \lambda}\)^2 &= {2 \a - F(r) \over G(r) \, H(r)}.
}} 
From the explicit functional 
behaviour of $F(r)$, $G(r)$, and $H(r)$, 
it is clear that $\C$ exists from any 
point with $ r > a^{1 \over d-4}$ as long as we make a judicious choice 
of $\a > 0 $. In particular,
since one can canonically choose $F(r) = 1 + c_4 \ln H(r)$, by choice of 
$ 2 \a \ge 1 + c_4 \ln \( H(r = {a^{1 \over d-4}} + \e)\)$, 
for arbitrarily small $\e$, we can 
achieve our desired objective of communicating to the asymptotic regions. 
For $c_4>0$, it suffices to let $\a > {1 \over 2}$.
Note that we can also use the same strategy as in \nobh\ to ``bound'' the 
spacetime \solng\ by a flat space metric, and use causality arguments
to show that there are no horizons.

One can write the metric \solng\ in a more suggestive form, by a change of 
coordinates. Let us denote $\alpha =  \sqrt{{2 (d-3) \over (d-2)}}$ and 
$\beta = {2 \over d-4}$. Then in dimensions $d \ge  5$ choosing 
$ r^{d-4} = a \, e^{2 x } $ we find that we can cast the metric in the form
\eqn\newcoords{
ds^2 = {1\over \( \tanh x \)^\a } \Big( -2 \, du \, dv + 
\( 1 + \ln \tanh x \) \, 
du^2 \Big) + \(\tanh x \)^{\alpha \beta} \(\sinh 2 x \)^\beta 
\Big( \beta^2 \, dx^2 + d \Omega^2 \Big). }
Here the coordinate $x \in (0,\infty)$ and the singularity is at 
$x =0$. For all values of $x > 0$, the metric is everywhere smooth,
and one doesn't expect there to be any horizons. 

In the four dimensional case writing, $\rho^2 = 1 + \ln r $, 
we obtain
\eqn\fourdnewcoods{
ds^2 = {1 \over \rho^4 } \( -2 \, du \, dv - \ln \rho^4 \, du^2 \) + 
16\, {1 \over \rho^8} \, d\rho^2 + \rho^2\, d\theta^2}
Clearly, this is not asymptotically flat. In fact, so much may be inferred 
from the original metric in \fourdim. This had to be the case, for in 
four dimensional spacetime, the transverse space being two-dimensional, 
causes the asymptotics to change quite drastically, leading to 
logarithmic deviations from flat space. We will not have much to say about 
this solution in what follows.

\subsec{Abandoning the $\dua$ isometry}

We now come to the interesting case, when $\dua$ is no longer an 
isometry. In this case, one might consider the situation wherein we 
allow all the functions appearing in \ansatz\ to have dependence on the 
$u$ coordinate as well. So we look for spacetimes with a metric ansatz
\eqn\notimeansatz{
ds^2 = {1 \over H(u,r)} \( -2 \, du \, dv + F(u,r) \, du^2 \)
      + G(u,r) \, (dr^2 + r^2 \, d\Om^2_{d-3}) + 2 A(u,r) \, du \, dr.
}
In fact, we can show that the last term, $2 A(u,r) \, du \, dr$, can 
be set to zero by appropriate coordinate transformation on $v$.
The simplest way to 
break the $\dua$ isometry is to just have the constants of integration in 
function $F(u,r)$ appearing in \solnF\ to be functions of $u$, {\it i.e.}, 
choose $ F(u,r) = f_1(u) + f_2(u) \ln H(r)$. The functions $H(r)$ and 
$G(r)$ are given as before in \solnH, \solnG. This is very similar to the 
fashion in which one breaks the analogous isometry in pp-waves.

The most general solutions to the metric ansatz \notimeansatz\ can also be 
found, and the functions appearing in the ansatz are given as:
\eqn\solnudep{\eqalign{
H(u,r) &= 
 h_1(u) \, \( {r^{d-4} -  h_2(u)\over r^{d-4} + h_2(u)} \)^{\sqrt{{2 (d-3) 
\over (d-2)}}} \cr
G(u,r) & = 
 g(u) \, \( {H(u,r)^2 \over \p_r H(u,r) \, r^{d-3} } \)^{2 \over d-4} \cr
F(u,r) & = f_1(u)  + f_2(u) \, \ln  H(u,r). 
}}
The solution in addition is 
characterized by four arbitrary functions $f_1(u)$, $f_2(u)$, $h_1(u)$
and $h_2(u)$ which are unconstrained. The function $g(u)$ is given in terms 
of $h_1(u)$ as  $g(u) = h_1(u)^{-{2 \over d-4}}$. 

Before ending this section, let us make a brief comment on the rotational
symmetry.  All the solutions presented above were by construction
rotationally symmetric in the transverse directions.  Since no nontrivial 
vacuum plane wave can preserve this symmetry, we could not have 
generated vacuum solutions which would be asymptotically plane wave.
(Note that this is no longer true if we relax the vacuum requirement.)
Indeed, as discovered above, all our solutions were asymptotically
flat, which is consistent, since flat spacetime is 
the only rotationally symmetric vacuum plane wave.
However, as we will demonstrate explicitly in the next section, 
we can deform these solutions in such 
a way as to break the rotational symmetry and at the same time lift the
asymptotic behavior to be that of a vacuum plane wave.

While this method generates the requisite asymptopia, it does not
modify the structure near the singularity sufficiently to generate
a horizon.  We will demonstrate this feature for an explicit example.
However, despite having broken the rotational symmetry, the metric
thus generated is still not the most general non-rotationally symmetric
spacetime.  In particular, one may ask whether one could not first
break the rotational symmetry so as to obtain an asymptotically 
flat solution with horizons, and then use the solution-generating
technique discussed below to make the solution have plane wave
asymptopia.  Whereas this is difficult to check analytically, 
we believe that such a scenario is unlikely, because of general 
properties of black holes.  
Specifically, one would expect that if a non-spherical solution exists,
we could obtain a spherically symmetric one by ``smearing'' or 
``superposing'' these solutions over all directions appropriately.
Intuitively, one might think of colliding the nonsymmetric black holes
so as to effectively restore the symmetry; and by the area theorem,
the final configuration should still possess a horizon.
Hence, the lack of such a symmetric black hole would naively suggest
the nonexistence of any non-spherically symmetric one.

\newsec{Deforming the solutions}

The vacuum solutions we wrote down in Section 2.2 are asymptotically flat. 
We are really interested in solutions which are 
asymptotically plane wave. Given an asymptotically flat solution 
we can use the Garfinkle-Vachaspati construction to deform it to be 
asymptotically plane wave.  We begin by reviewing the construction 
in a more general (non-vacuum) setup,
and then proceed to apply the same to the vacuum solutions discussed in 
Section 2.

\subsec{Review of the Garfinkle-Vachaspati construction}

The Garfinkle-Vachaspati (GV) construction \gava, \garfinkle, is 
essentially a solution generating technique. Given a solution 
to Einstein's equations (in general, with some appropriate 
matter content), with certain specific properties, one 
can deform the solution to a new one with the same matter
fields. In particular, the scalar curvature invariants of the deformed 
solution are identical to the parent solution. The  
idea is that given a solution with an appropriate set of symmetries,
one can essentially ``linearize'' Einstein's equations, which will then 
allow one to superpose solutions. This technique was developed in an 
attempt to add wavy hair to black holes/strings. We will in the 
following briefly review the construction as presented in \kmr.

We assume that we are working with the Einstein-Hilbert action with 
some matter fields. To be specific, we assume that the matter 
content is the conventional matter 
appropriate for supergravity theories. In particular, we will have
scalar fields $\phi_i$, and a bunch of  $p-$form fields $A_p$ with their 
associated field strengths $F_{(p+1)}$. We will work with the action,
\eqn\einshilact{
S = \int \, d^d x\, \sqrt{-g} \, \( R - {1 \over 2} \, \sum_{i} \,
 \alpha_i(\phi) \, \(\nabla \phi_i \)^2 - { 1\over 2} \, 
\sum_{p} \, \beta_p(\phi) \, F_{(p+1)}^2 \)
}
Thus, the metric we work with will always be the Einstein frame metric, 
although one might generalize the discussion to work with the string 
frame metric. The couplings $\alpha_i(\phi)$ and $\beta_p(\phi)$ 
are assumed to be non-derivative couplings for convenience.

The primary requirement for a solution,
$\CM(g_{\mu\nu}, \phi_i, F_{(p+1)})$, to be amenable to the 
GV construction is that it possess a null, hypersurface-orthogonal 
Killing field {\it i.e.}, the solution admits a 
vector field $k^\mu$ which satisfies
\eqn\gvreq{\eqalign{
k_\mu \, k^\mu & = 0  \cr
\nabla_{(\mu} k_{\nu )} &= 0 \cr
\nabla_{[\mu} k_{\nu]} & = k_{[\mu} \nabla_{\nu]} S
}}
The first two conditions demand that the vector be null and Killing, 
while the last condition is that of hypersurface-orthogonality. $S$ is 
a scalar function on the background geometry. Note that if $S$ is a 
constant, then the vector $k^\mu$ must be covariantly constant. 
This is the case for pp-waves.

With respect to the matter fields supporting the metric in 
$\CM(g_{\mu\nu},\, \phi_i, \,F_{(p+1)})$, we require that the 
scalar fields have a vanishing Lie derivative with respect to the 
vector $k^\mu$ {\it i.e.}, ${\cal {L}}_k \, \phi_i = 0$. We impose 
the same constraint on the $(p+1)-$form field strengths; 
${\cal{L}}_k \, F_{(p+1)} = 0$. However, it will turn out that in 
order to maintain the form of the field strength in the deformed 
solution, we will require further that they satisfy an additional 
transversality condition $ i_k \, F_{(p+1)} = \omega_k \, \wedge \, 
\theta_{(p-1)}$. Here, $i_k$ denotes the interior product, while 
$\omega_k$ is the one-form associated with the vector $k^{\mu}$, 
and $\theta_{(p-1)}$ is some $(p-1)-$form. These requirements are 
necessitated to ensure that upon deforming the metric, the change in 
stress tensor, given the functional form of the matter fields in 
$\CM(g_{\mu\nu}, \, \phi_i, \,  F_{(p+1)})$,
is such that one is left with a linear equation for the 
deformation parameter.

Given $\CM(g_{\mu\nu}, \, \phi_i, \, F_{(p+1)})$ as a solution to 
\einshilact\ with the above properties, we can find a new solution 
$\hat{\CM}(G_{\mu\nu}, \, \phi_i, \, F_{(p+1)})$, where the new 
metric $G_{\mu \nu}$ is defined in terms of a deformation $\Psi$ 
as follows:
\eqn\gvdeform{
G_{\mu \nu} = g_{\mu \nu} + e^S \, \Psi \, k_\mu \, k_\nu.
}
As the notation suggests, the form of the matter fields is left 
unchanged. The function $\Psi$ satisfies 
the following constraints
\eqn\psiconstr{
k^\mu \, \nabla_\mu \, \Psi = 0, \;\;\;\;\;\; {\rm and} \;\;\;\;\;\;\;
\nabla^2 \, \Psi =0.
}
The first of these is equivalent to demanding that $\Psi$ has vanishing 
Lie derivative along $k^{\mu}$, and the second just demands that 
$\Psi$ solve the covariant Laplace equation in the original background, 
{\it i.e.}, the Laplacian with respect to the metric $g_{\mu \nu}$.
It may be verified that the deformed solution retains the 
scalar curvature invariants of the parent solution. 
For further details of the construction and proof of the equivalence 
of the scalar curvature invariants the reader 
is referred to 
{\it e.g.}\ \kmr.

As a simple example of the GV construction, consider a plane wave 
spacetime, with the metric 
\eqn\plane{
ds^2 = - 2 \, du \, dv - f_{ij}(u) \, x^i \, x^j  \, du^2 + (dx^i)^2.
}
We assume that the metric is supported by an appropriate stress tensor, 
which goes into the determination of the precise form of the functions 
$f_{ij}(u)$. Now, the Einstein's equations are linear 
because of the fact that the background admits a covariantly 
constant null Killing field $\dva$. Consider now the metric
\eqn\planedef{
ds^2 =  - 2 \, du \, dv - \( f_{ij}(u) \, x^i \, x^j  - F(u,x^i) \)
\, du^2 + (dx^i)^2.}
This metric will satisfy the equations of motion with the same matter 
content, so long as $F(u,x^i)$ is harmonic in the transverse dimensions
$x^i$.
(In terms of \gvdeform, $S \equiv 0$ and $\Psi \equiv F$.)
But \planedef\ has the general form of a pp-wave spacetime. Thus, 
one trivial application of the GV construction is to deform any 
plane wave spacetime into a pp-wave spacetime. 
 
\subsec{Vacuum solutions and Garfinkle-Vachaspati construction}

In the present subsection we will apply the GV construction to the 
class of solutions presented in section 2. In this case,
since we have no matter fields present, the discussion is much simplified. 
All we need to do is to ensure that the metric satisfies the appropriate 
constraints as in \gvreq\ and to find suitable deformations of the same.

Let us start with the form of the metric as in \ansatz, wherein for 
simplicity of discussion we will revert back the case with the extra 
 isometry. As mentioned earlier, it is 
clear that $\dva$ is a null Killing field. To check that it in addition is 
hypersurface-orthogonal is trivial and one finds that $S = \ln \, H(r)$. 
So the new metric according to \gvdeform\ will in general be of the form 
\eqn\gvansatz{
ds^2 = {1 \over H(r)} \( -2 \, du \, dv + F(r) \, du^2 \)
      + G(r) \, (dr^2 + r^2 \, d\Om^2_{d-3}) + {1 \over H(r)} \, 
\Psi(u,v,r,\Om_{d-3})
\, du^2.}

The first of the conditions in \psiconstr\ implies that  
$\partial_v \, \Psi(u,v,r,\Om_{d-3}) =0 $ and therefore 
the second implies that 
$\Psi(u,r,\Om_{d-3})$ satisfy:
\eqn\vacharmonic{
\nabla^2 \, \Psi = {H(r) \over r^{d-3} \, G(r)^{{d-2 \over 2}}} 
\, {d \over dr} \(
 { G(r)^{{d-4 \over 2}} \, r^{d-3} \over H(r) }\,  {d \Psi \over dr} \) + 
{1 \over r^2 G(r)} \,  \nabla_{\Om_{d-3}}^2 \, \Psi =0} 
Decomposing $\Psi(u,r,\Om_{d-3})$ into spherical harmonics on the 
${\bf S}^{d-3}$, labeled by $L = (\l,m_1, \cdots)$, 
with principal angular momentum $\l$, the general 
solution takes the form:
\eqn\vacpsi{
 \Psi(u,r,\Om_{d-3}) = \sum_{L} \, \xi_L(u) \, \psi_{\l}(r) \, Y_L(\Om_{d-3}).
}
The function $\psi_{\l}(r)$ satisfies a one-dimensional radial wave equation
\eqn\psivacrad{
{H(r) \over r^{d-3} \, G(r)^{{d-4 \over 2}}} \, {d \over dr} \(
 { G(r)^{{d-4 \over 2}} \, r^{d-3} \over H(r) }\,  
{d \psi_{\l}(r) \over dr} \) - 
{\l (\l +d-4) \over r^2} \, \psi_{\l}(r) =0.}
Solutions of this equation will determine the generically allowed 
deformations of our vacuum solutions. 
Using the relation between the functions $G(r)$ and $H(r)$ as in 
\solnG, this can be simplified to read
\eqn\psivacrad{
{H'(r) \over H(r)} \, {d \over dr} \(
 { H(r)\over H'(r) }\,  
{d \psi_{\l}(r) \over dr} \) - 
{\l (\l +d-4) \over r^2} \, \psi_{\l}(r) =0.}

However, we are interested in the particular case when asymptotically 
the solution is of the plane wave form. To determine if this is 
possible, all 
we need is to examine the asymptotics of \psivacrad\ and demand that the 
solution behave like $r^2$.  In general, given that the 
functions $H(r)$ and $G(r)$ tend to one as $r \rightarrow \infty$, 
in dimensions $d \ge 5$, we 
see that the asymptotics of the solution for $\Psi$ is the  same as 
a harmonic function in $(d-2)$ dimensions. So we are guaranteed to have a 
solution wherein $\Psi(u,r \to \infty,\Om_{d-3}) \rightarrow r^2 \, 
Y_2 \( \Om_{d-3} \)$. We therefore need to solve the 
radial equation \psivacrad\ for the $\l = 2 $ mode in order to 
find solutions that are asymptotically plane wave. 

As an example let us consider the vacuum solution \solng\ in five dimensions.
The resulting metric is explicitly given as 
\eqn\fivedsoln{\eqalign{
ds^2 &= {1 \over H(r)} \Bigg( -2 \, du \, dv + \[
1+ \ln H(r) +  \xi_2(u)\, 
\psi_2(r) \(3 \cos^2\theta -1 \) \]
 \, du^2 \Bigg) \cr 
&\;\;\;\;\;\;\;\;\;\;\;\;\;
     + {H(r)^4 \over r^4 \, H'(r)^2  } \, \Bigg( dr^2 + r^2 \, d\theta^2 + 
r^2 \,\sin^2\theta \, d\phi^2
\Bigg) \cr
H(r) & = \( {r -1 \over r+1} \)^{{2\over \sqrt{3}}} \cr
\psi_2(r) & =  \( 3 r^2 + 2 + {3 \over r^2} \) \, \( \a_1 + \a_2 
\, \ln \({r-1 \over r+1} \) \) + 6 \a_2 \, \(r + {1 \over r}\)
}}
with arbitrary integration constants $\a_1$ and $\a_2$ and as before
$\xi_2(u)$ is an arbitrary function of $u$. The 
asymptotic behaviour of the solution is 
clearly that of a vacuum plane wave in five dimensions, $\CV_5$.

\newsec{Supersymmetric solutions}
In Section 2, we saw that vacuum solutions to Einstein's equations
with a globally null Killing field and transverse rotational 
invariance, are uninteresting from the point of view of being 
black string solutions with a regular horizon. Nonetheless, these 
can be deformed to be asymptotically plane wave, notwithstanding the 
fact that we are still eluded from our goal of finding  
asymptotically plane wave black strings. One might 
be tempted to claim that the requirement of a global null symmetry is 
too restrictive to allow for black holes/strings. However, the fallacy 
of this logic is all too apparent when one considers solutions to Einstein's
equations with appropriate matter content. In particular, the solution 
presented in \nobh\ did possess a globally null Killing field and 
was a solution to the low energy effective action of 
the five dimensional heterotic string.

In supergravity theories we can easily find many examples of 
solutions which  admit a globally null Killing field and in addition have 
regular horizons. To illustrate this point, it is worthwhile to consider the 
most famous supersymmetric solutions, the extremal black branes in 
ten (IIA/IIB) or eleven dimensional supergravity.
In eleven dimensional supergravity we have the extremal branes,
and in IIA (IIB) supergravity we have the even (odd) Dp-brane 
solutions, all of which are asymptotically flat.
Since the extremal black brane solutions have an 
isometry group $SO(1,p) \times SO(9-p)$, for $p > 1$ it is clear 
that we have always two null isometries in the solution. 
Given two everywhere null Killing fields, we can deform the 
solution using the GV construction and make it asymptotically plane wave.

\ifig\figbranes{Penrose diagrams of the extremal black branes.
(a) corresponds to the M5-brane and D3-brane,
(b) corresponds to the M2-brane, and (c) denotes the other Dp-branes for 
$p < 6$ and $p \neq 3$.}
{\epsfxsize=12cm \epsfysize=9cm \epsfbox{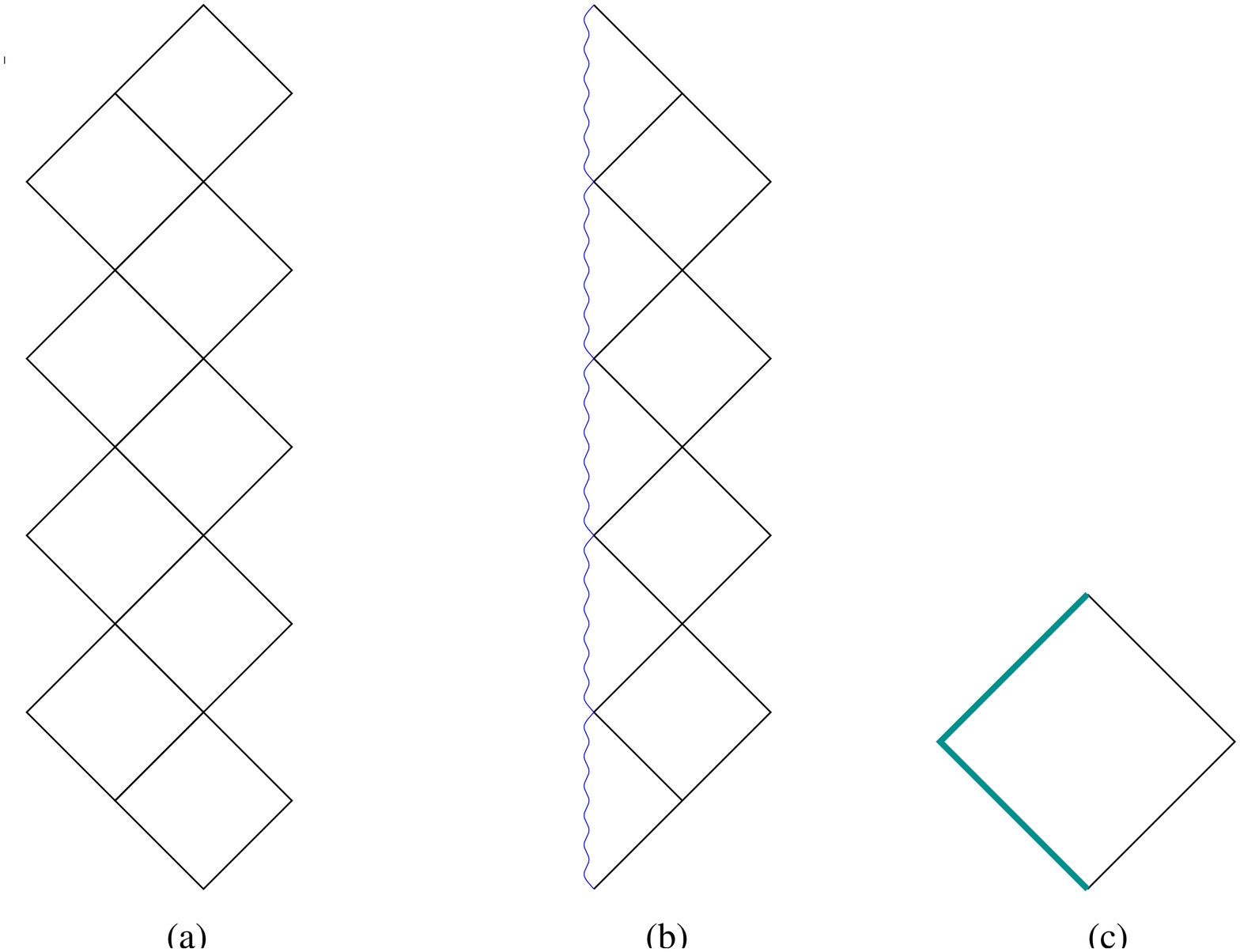}}

Of these solutions, the most interesting are 
the D3, M2 and M5 branes. The extremal D3 and the M5 brane 
solutions are asymptotically flat and admit a maximally extended spacetime
which is everywhere smooth and has regular horizons. The near-horizon 
geometry is of course the well known $AdS_5 \times {\bf S}^5$ spacetime
for D3 branes and $AdS_7 \times {\bf S}^4$ for the M5 brane. 
In the case of the M2-brane, the horizon cloaks a timelike singularity, 
and the causal structure of the maximally-extended 
spacetime looks like that of 
the extremal Reissner-Nordstr{\" o}m black hole in four dimensions. 
The near horizon geometry is of course $ AdS_4 \times {\bf S}^7$. 
The remaining D$p$-brane solutions for $p \le 6$
have the singularity coincident with the
horizon and in this sense do not posses a regular horizon. The 
Penrose diagrams for the various brane solutions are given in Fig.1. 

Thus, while all the D-branes are in principle candidates for deformation 
into an asymptotically plane wave geometry using the GV construction, 
only the M2-brane solution is capable of producing a spacetime which 
has in addition a regular event horizon cloaking a singularity.
While the GV construction does not guarantee that the causal 
structure of the original and the deformed spacetimes will remain the
same, the curvature invariants of the two metrics are identical.
Therefore, for general Dp-brane solutions, the deformation by the 
GV construction to make the spacetime asymptotically plane wave
will not alleviate the singular property of the horizon.
For the case of the non-dilatonic branes with a regular event horizon,
we will see that the term added to implement the GV construction
will be vanishingly small in the vicinity of the horizon, 
thereby preventing the horizon from being destabilized in the resulting
spacetime (for an explicit check, see the analysis in \kmr\ for the
5-dimensional black hole case).
So in what follows, we shall dwell mostly on the case of the M2-brane 
and only briefly mention the other cases.

\subsec{Deforming the M2-brane}

In the following we show that it is possible to deform the M2-brane 
solution in eleven dimensional supergravity so that the asymptotic
behaviour is $\CV_{10} \times \R$, whilst retaining the nature of the  
near-horizon geometry and the singularity intact. For a 
discussion of supergravity solutions of Dp-branes in plane wave backgrounds, 
which give rise to similar metrics see \bmz. 

The M2-brane solution is given as 
\eqn\mtwosol{\eqalign{
ds^2 &= H_2(r)^{-{2 \over 3}} \, \( -dt^2 + dx_1^2 + dx_2^2 \) + H_2(r)^{{1\over 3}}
\, \( dr^2 + r^2 \, d\Om_7^2 \) \cr
F_4 &= \({d H_2(r)^{-1} \over dr }\) \, dt \wedge dx_1 \wedge dx_2 \wedge dr,
}}
where $H_2(r) = 1 + {Q^6 \over r^6}$ is a harmonic function in the transverse 
eight-dimensional space. The horizon is at $r =0$ in these coordinates. 
To see the location of the singularity, it is best to define a new coordinate
$\zeta = r^2$. The singularity is located at the zero of $H(\zeta)$, 
{\it i.e.}, $ \zeta = - Q^2$.

We can define two null directions $u,v$ by linear combinations of $t$ and 
$x_1$ and write the solution as 
\eqn\mtwosoluv{\eqalign{
ds^2 &= H_2(r)^{-{2 \over 3}} \, \( -2 \, du \, dv+ dx_2^2 \) + H_2(r)^{{1\over 3}}
\, \( dr^2 + r^2 \, d\Om_7^2 \) \cr
F_4 &= \({d H_2(r)^{-1} \over dr }\) \, du \wedge dv \wedge dx_2 \wedge dr.
}}
Now, both $\dva$ and $\dua$ are null Killing vectors and we can also 
verify that they are hypersurface-orthogonal. Moreover, the field 
strength satisfies the requisite transversality condition. In short, we 
are allowed to deform the solution using the GV construction by choosing 
one of the two null Killing vectors. To wit, we can write the deformed metric 
as 
\eqn\gvmtwodef{
ds^2 = H_2(r)^{-{2 \over 3}} \, \( -2 \, du \, dv+ dx_2^2  - 
\Psi(u,x_2,r,\Om_7) \, du^2
\) + H_2(r)^{{1\over 3}}
\, \( dr^2 + r^2 \, d\Om_7^2 \) 
}
We have already taken into account the first of the constraints on $\Psi$
in \psiconstr, by having $\partial_v \, \Psi =0$. The second constraint 
reduces to 
\eqn\psieq{
H_2^{1/3}(r) \,
\nabla^2 \, \Psi = H_2(r) {d^2 \Psi \over dx_2^2}  + {d^2 \Psi \over dr^2 } +
{7 \over r } \, {d\Psi \over dr} + {1\over r^2} \nabla_{\Om_7}^2 \, \Psi =0
}
Writing $\Psi(u,x_2,r,\Om_7) = \xi_{kL}(u) 
\, e^{i k x_2} \, \psi_{k \l}(r) \, 
Y_L(\Om_7) $ with arbitrary functions $\xi_{kL}(u)$, 
and $Y_L(\Om_7)$ denoting the spherical harmonics with $L$ being a 
label for the set of angular momenta on the seven sphere with principal 
angular momentum $\l$, we find the radial equation
\eqn\psimtworad{
 {d^2 \psi_{k \l}(r) \over dr^2 } +
{7 \over r } \, {d\psi_{k \l}(r) \over dr} - \({\l(\l+6) \over r^2} + 
k^2 \, H_2(r) \)\, \psi_{k \l}(r)   =0
}

For the case of $ k =0$, {\it i.e.}, requiring $\( \partial \over \partial x_2
\)^a$ to be a Killing vector in the deformed geometry, we are 
reduced to solving the Laplace equation in eight dimensional flat space.
So we can pick the $\l =2 $ mode on the seven sphere to obtain a 
solution which is asymptotically plane wave.
 
Let us parametrize the seven sphere by the coordinates 
such that $\theta$ corresponds to the azimuthal angle, and choose
$\Psi(u,x_2,r,\Om_7) = r^2 \, \(8 \cos^2 \theta -1 \)$. Then we claim that 
\eqn\mtwodefsol{\eqalign{
ds^2 &= H_2(r)^{-{2 \over 3}} \, \[ -2 \, du \, dv+ dx_2^2  - 
r^2  \( 8 \, \cos^2\theta -1 \) \,du^2
\] \cr 
& \;\;\;\;\;\;\;\; + H_2(r)^{{1\over 3}}
\, \[ dr^2 + r^2 \( d\theta^2 + \sin^2\theta \, d\Om_6^2 \) \] \cr
F_4 &= \({d H_2(r)^{-1} \over dr }\) \, dt \wedge dx_1 \wedge dx_2 \wedge dr,
}}
is a solution to eleven dimensional supergravity, with the same 
curvature invariants as the M2-brane solution. In particular, the solution 
still has a regular horizon at $r =0$ 
(one can also check the regularity of the horizon by looking at
the near-horizon geometry). This is to be contrasted with 
added pp-wave like terms $\Psi(u,r) \sim {1 \over r}$, or 
having plane waves along the longitudinal directions of the 
brane, wherein one does encounter singularities at the horizon \bcr\ 
({\it cf.}\ \refs{\mone, \mtwo, \kmr, \TseytlinA, \TseytlinB, 
\gy, \myers, \ross} for discussions of similar issues 
in the case of black strings/branes).
Its asymptotic behaviour is that 
of a ten dimensional vacuum plane wave times an extra real line parameterized 
by $x_2$. Since the curvature invariants in \mtwodefsol\ are identical to
those in the parent M2-brane solution \mtwosol, we see that the 
singularity at $\zeta = -Q^2$ remains unchanged.

In the event of $k \neq 0$ we see that the nature of the asymptotics 
can be changed quite dramatically. The asymptotic behaviour of the 
equation \psimtworad\ can be analyzed to show that the solutions are 
Bessel functions which either diverge or decay faster than $r^2$.
So by looking for solutions wherein we have some momentum along the brane 
world-volume directions we do not obtain a solution that looks like a 
plane-wave. Thus, in order to obtain a plane wave solution, our 
only choice is to use the solution presented in \mtwodefsol, wherein 
we have an additional isometry corresponding to translations along 
$\( \partial \over \partial x_2\)^a$. 

It is tempting to ask whether it is not possible  to deform the 
M2-brane solution, so that the asymptotic solution is that of the 
maximally supersymmetric plane wave in IIB supergravity (BMN plane wave).
To achieve this we want the function $\Psi(u,x_2,r,\Om_7) = r^2$. 
Let us therefore consider the following metric:
\eqn\mtwobmn{
ds^2 = H_2(r)^{-{2 \over 3}} \, \[ -2 \, du \, dv+ dx_2^2  - 
r^2  \,du^2
\] + H_2(r)^{{1\over 3}}
\, \[ dr^2 + r^2 \( d\theta^2 + \sin^2\theta \, d\Om_6^2 \) \]}
Given this metric one can readily calculate the stress tensor $T_{\mu \nu}$
supporting the solution. Writing $T_{\mu \nu} = T^{(F_4)}_{\mu \nu} + 
\tilde{T}_{\mu \nu}$, with $T^{(F_4)}_{\mu \nu}$ being the contribution to the 
stress tensor from the background 4-form field strength, 
$F_4 =   \({d H_2(r)^{-1} \over dr }\) \, dt \wedge dx_1 \wedge dx_2 \wedge dr$, and $\tilde{T}_{\mu \nu}$ is the additional part of the stress tensor 
necessary to satisfy Einstein's equations. One finds that the only 
non-vanishing component of $\tilde{T}_{\mu \nu}$ is given as 
\eqn\stress{
\tilde{T}_{uu} = { 8 \over H_2(r)}
}
It is not possible to  generate the additional piece of the stress tensor 
using the 4-form field strength. While one can readily write down 
non-vanishing components of the 4-form to generate \stress, for instance by 
writing $F_{u x_2 r \theta} = {4 \sqrt{6} r \over H_2(r)}$, 
it is not possible to satisfy the field strength equations of motion. So 
we conclude that there is no consistent solution to the equations of motion 
of 11-dimensional supergravity with the metric given as in \mtwobmn. One 
can however make \mtwobmn\ a solution if one had, in addition to 
the field strength, some null dust with stress tensor \stress. 

\subsec{The other black branes}

Let us now turn to the other extremal brane solutions. In eleven 
dimensional supergravity, we can consider the M5-brane solution, given by
\eqn\mfivesol{
ds^2 = H_5(r)^{-{1 \over 3}} \, \( -dt^2 + dx_1^2 + dx_2^2 +dx_3^2 + 
dx_4^2 + dx_5^2 \) 
+ H_5(r)^{{2\over 3}}
\, \( dr^2 + r^2 d\Om_4^2 \) 
}
with $H_5(r) = 1+{ Q_5^3 \over r^3}$ and the field strength 
$F_4 \propto \omega_4$, the volume form on the transverse ${\bf S}^4$.
Now, we can again combine $t$ and any one of the longitudinal directions 
of the M5-brane, say $x_1$ to form two null directions, $u,v$. Repeating the 
steps as in the previous example with the M2-brane, we see that the 
following is a solution to eleven dimensional supergravity:
\eqn\mfivedefsol{\eqalign{
ds^2 &= H_5(r)^{-{1 \over 3}} \, \[ -2\, du \, dv  + \sum_{i=2}^5 
\, (dx_i)^2 - r^2 \,\(5 \, \cos^2 \theta -1 \) \, du^2 \] \cr
& \;\;\;\;\;\;\;\; + H_2(r)^{{2\over 3}}
\, \[ dr^2 + r^2 \( d\theta^2 + \sin^2\theta \, d\Om_3^2 \) \]. 
}}
The four-form field strength supporting the solution remains 
proportional to the volume form on the ${\bf S}^4$. This solution 
interpolates between a near-horizon geometry ($r \ll 1$) of 
$AdS_7 \times {\bf S}^4$ and a seven dimensional vacuum plane 
wave with four flat longitudinal directions parameterized by $x_i$, 
$i =2, \cdots , 4$. 

We can play the same game with the extremal brane solutions to 
ten dimensional supergravity. 
The extremal D$p$-brane solutions, in the string frame metric, 
are given by
\eqn\dpsoln{\eqalign{
ds_{str}^2 & = H_p(r)^{-{1\over 2}} \, \( -dt^2 + 
\sum_{i=1}^p \, (dx^i)^2\) 
+ H_p(r)^{{1\over 2}} \, \(dr^2 + r^2\, d\Om_{8-p}^2 \) \cr
e^{4 \Phi}  & = H_{p}(r)^{3-p} \cr
F_{(p+2)} & = \({d H_p^{-1}(r) \over dr} \) \, dt \wedge 
dx^1 \wedge \cdots dx^p \wedge dr
}}
Yet again defining $ u$, $v$ as linear combinations of $t$ and $x^1$, we 
have two null directions which are conducive to applying the GV technique.
Note that in addition to the form fields we also have a non-vanishing 
dilaton in the cases when $p \neq 3$. The dilaton, when non-vanishing,
does satisfy the requirement that it have vanishing Lie derivative along the 
null Killing vector. 

By applying the GV technique, we find new solutions to the supergravity 
equations of motion;
\eqn\dpsoln{\eqalign{
ds_{str}^2 & = H_p(r)^{-{1\over 2}} \, \[ -2\, du \, dv  - \Psi(u,x^i, 
r, \Om_{8-p}) \, du^2 +   
\sum_{i=2}^p \, (dx^i)^2\] 
+ H_p(r)^{{1\over 2}} \, \(dr^2 + r^2\, d\Om_{8-p}^2 \) \cr
e^{4 \Phi}  & = H_{p}(r)^{3-p} \cr
F_{(p+2)} & = \({d H_p^{-1}(r) \over dr} \) \, dt \wedge 
dx^1 \wedge \cdots dx^p \wedge dr
}}
with $\Psi(u,x^i,r,\Om_{8-p}) = 
\xi_{kL}(u)
\, \exp\( i \sum_{i =2}^p \, k_i x_i \) \, \psi_{k \l}(r)  \,
Y_L(\Om_{8-p}) $. Here $\xi_{kL}(u)$ are arbitrary functions of $u$, with 
$k = \{k_2, \cdots, k_p\}$, and $Y_L(\Om_{8-p})$ are the spherical harmonics 
on the $\S^{8-p}$ labeled by $L = ( \l , m_1, \cdots)$. $\psi_{k \l}(r)$
satisfies the radial wave equation
\eqn\dppsi{
{d^2 \psi_{k \l}(r) \over dr^2 } +
{(8-p) \over r } \, {d\psi_{k \l}(r) \over dr} - \({\l(\l+7-p) \over r^2} + 
\sum_{i = 2}^p \, (k_i)^2 \, H_p(r) \)\, \psi_{k \l}(r)   =0
}
It is clear that in the case $k_i \equiv 0$, one has solutions that behave 
like $r^2$ asymptotically, for $l =2$. Thus, it is possible to 
deform the D$p-$brane solutions so that their asymptotic behaviour 
changes from $\R^{1,9}$ to $\CV_{11-p} \times \R^{p-1}$. However, as argued 
earlier, D$p-$brane solutions with $p \neq 3$ do not admit a regular horizon.
In the case of the D3-brane the GV construction provides an example of a 
solution that smoothly interpolates between a 
near-horizon geometry of  $AdS_5 \times \S^5$ and an  
asymptotic behaviour of $\CV_8 \times \R^2$.

\newsec{Discussion}

We have discussed solutions to vacuum Einstein's equations with a globally 
null Killing field and a transverse rotational symmetry. 
Knowledge of the explicit form of the solutions enabled us to 
conclude that there are no horizons in these spacetimes. While
one would have thought that, in order to find vacuum asymptotically  plane 
wave solutions, it would have been necessary to break the transverse 
rotational symmetry from the outset, we demonstrate that it is possible to 
deform the rotationally symmetric ones into the desired form.

However, the main motivation for undertaking the exercise hasn't been 
realized, in the sense that we haven't found a vacuum black hole solution that 
is asymptotically plane wave. It is clear that the requirement of a 
globally null Killing field is too restrictive to allow for black hole 
solutions in vacuum gravity. On the other hand, in theories with 
some matter content we are able to generate the solutions possessing 
regular event horizons and having the desired asymptotics.
Before turning our attention to these, we note that
a possible strategy at generating vacuum solutions which have horizons and 
the correct asymptotics is to abandon the requirement of a globally null 
Killing field, by considering a small $g_{vv}$ 
contribution to the metric. One such possibility is to  
attempt to patch a small black hole 
solution into the vacuum plane wave geometry. 
However, we leave these directions for future investigation.

The situation in terms of charged black holes is much better. Here we have 
shown that it is possible to generate a large class of new solutions 
from the previously known ones. One can 
use the recent results on the supersymmetric solutions 
of minimal supergravity in five dimensions \gaunt\ to show that it is 
possible to have solutions which are asymptotically plane wave and 
yet admit a regular horizon in certain cases.
In this context, the solutions generated from 
M2-branes are perhaps the most interesting, from the point of view of 
generating asymptotically plane wave black holes. On the other hand, the 
solutions obtained by deforming the extremal M2, M5 and D3-brane metrics 
are extremely interesting from the perspective of the holographic 
dual, for the deformation of the spacetime geometry ought to correspond
to some particular deformation of the field theory.

One can illustrate this with the situation of the D3-brane.
Consider starting from the near-horizon geometry of the D3-brane, {\it i.e.}, 
$AdS_5 \times \S^5$. From the AdS/CFT duality we know that this is 
related to the $\CN =4$, $d =4$ Super-Yang-Mills theory. In the dual gauge 
theory, turning on an irrelevant operator, ${\rm Tr} \, F^4$, 
is believed to correspond to reconnecting the asymptotically flat 
region in the extremal D3-brane geometry, with the near-horizon 
$AdS_5 \times \S^5$ region, by recreating the throat geometry that 
interpolates between the two \ghk, \gubser. ({\it cf.} \costa, \costab, \dgks, 
and \ejp, for related discussions).

We have shown that there is a deformation in the full extremal
D3-brane geometry \dpsoln\ (for $p=3$), which changes the asymptotically 
flat region $\R^{9,1}$ into $\CV_8 \times \R^2$. The resulting 
geometry is 
\eqn\threedef{\eqalign{
ds^2 &= \( 1 + {R^4 \over r^4} \)^{-{1\over 2}} \, \[ -2 \, du \, dv - r^2 \, 
\( 6 \, \cos^2 \theta -1 \) \, du^2 + \sum_{i =2}^3 \, \( d x^i \)^2 \] + \cr 
& \hskip8mm
\(1 + {R^4 \over r^4}\)^{{1\over 2}} \, 
\[ dr^2 + r^2 \(  d\theta^2 + \sin^2 \theta \,
d\Om_4^2 \) \].
}}
The geometry in the near-horizon limit $r \ll 1$, reduces to 
$AdS_5 \times \S^5$ with a deformation proportional to $r^2$. In the 
dual $\CN =4$ gauge theory this would correspond to a 
deformation by a dimension 6 operator such as ${\rm Tr} \, F^3$. One can 
envisage starting from the $\CN =4$ gauge theory and deforming it first 
by the dimension eight operator ${\rm Tr} \, F^4$, so as to recover the 
asymptotically flat region and then further deform it by an 
operator that had dimension six in the $AdS$ limit so as
to reproduce the supergravity geometry 
as in \threedef. One might substantiate the claim further by noting that 
the full extremal D3-brane geometry is holographically related to 
non-commutative Super-Yang-Mills theory with a self-dual B-field \dast.
One can then ask what is the corresponding deformation in the 
non-commutative gauge theory which reproduces the geometry \threedef? 
This issue deserves to be investigated further. Similar statements 
can be made for the M2 and M5 brane geometries, but one is handicapped 
by not having a precise understanding of the dual field theories.


\vskip 1cm

\centerline{\bf Acknowledgements}
It is a great pleasure to thank Ori Ganor and 
Gary Horowitz for  discussions, and Gary Horowitz for comments 
on the manuscript.
VH was supported by NSF Grant PHY-9870115, while MR acknowledges support 
from the Berkeley Center for Theoretical Physics and also partial support 
from the DOE grant DE-AC03-76SF00098 and the NSF grant PHY-0098840.

%

\listrefs
\end